\documentclass[conference]{IEEEtran}
\IEEEoverridecommandlockouts
\usepackage{cite}
\usepackage{amsmath,amssymb,amsfonts}
\usepackage{algorithmic}
\usepackage{graphicx,import}
\graphicspath{ {figures/} }
\usepackage{textcomp}
\usepackage{xcolor}
\usepackage{float}
\usepackage{stfloats}
\usepackage{Robertocommands}
\def\BibTeX{{\rm B\kern-.05em{\sc i\kern-.025em b}\kern-.08em
    T\kern-.1667em\lower.7ex\hbox{E}\kern-.125emX}}

\begin{document}

\title{Sensing Micro-colloid Concentration by Spectral Impedance Measurements and Relaxation Times Analysis \\
%\thanks{Identify applicable funding agency here. If none, delete this.}%
}

%\title{Sensing Micro-colloid Concentration by Impedance Measurements and Time Constant Spectroscopy \\
%\thanks{Identify applicable funding agency here. If none, delete this.}
%}

\author{\IEEEauthorblockN{Roberto G. Ram\'irez-Chavarr\'ia}
\IEEEauthorblockA{\textit{Instituto de Ingenier\'ia,} \\ \textit{Instituto de Ciencias Aplicadas y Tecnolog\'ia,}\\
\textit{Universidad Nacional Aut\'onoma de M\'exico}\\
Mexico City, 04510, Mexico \\
roberto.ramirez@icat.unam.mx}
\and
\IEEEauthorblockN{Celia S\'anchez-P\'erez}
\IEEEauthorblockA{\textit{Instituto de Ciencias Aplicadas y Tecnolog\'ia} \\
\textit{Universidad Nacional Aut\'onoma de M\'exico}\\
Mexico City, 04510, Mexico \\
celia.sanchez@icat.unam.mx}
%\and
%\IEEEauthorblockN{3\textsuperscript{rd} Given Name Surname}
%\IEEEauthorblockA{\textit{dept. name of organization (of Aff.)} \\
%\textit{name of organization (of Aff.)}\\
%City, Country \\
%email address}
%\and
%\IEEEauthorblockN{4\textsuperscript{th} Given Name Surname}
%\IEEEauthorblockA{\textit{dept. name of organization (of Aff.)} \\
%\textit{name of organization (of Aff.)}\\
%City, Country \\
%email address}
%\and
%\IEEEauthorblockN{5\textsuperscript{th} Given Name Surname}
%\IEEEauthorblockA{\textit{dept. name of organization (of Aff.)} \\
%\textit{name of organization (of Aff.)}\\
%City, Country \\
%email address}
%\and
%\IEEEauthorblockN{6\textsuperscript{th} Given Name Surname}
%\IEEEauthorblockA{\textit{dept. name of organization (of Aff.)} \\
%\textit{name of organization (of Aff.)}\\
%City, Country \\
%email address}
}

\maketitle

\begin{abstract}
This paper presents an attractive method towards sensing and quantifying the concentration of dielectric particles suspended in a highly conductive medium. Fast electrical impedance spectroscopy (EIS) measurements were performed, based on an electrochemical setup and a non-parametric unbiased estimator. The impedance data are then mapped from the frequency domain to a time constant domain via the distribution of relaxation times (DRT) model. It is shown that application of DRT can systematically give a distribution function which is directly related with the concentration of micro-colloid samples. As a result, the samples were characterized with high sensitivity and resolution.
\end{abstract}
\vskip 0.2cm
\begin{IEEEkeywords}
Micro-colloid, Electrical Impedance Spectroscopy, Distribution of Relaxation Times, Spectroscopy, Impedance Sensor
\end{IEEEkeywords}

\section{Introduction}
Electrical properties of colloidal samples, particularly suspensions of dielectric particles embedded on a conductive medium, have been studied for many years~\cite{Asami2002}. Moreover, their dielectric and conductive behavior is known to be related with their structure and composition.
When a colloidal suspension of dielectric particles is exposed to an external electric field, they become polarized, accumulating charge at the interface of its shell and the surrounding medium.
The degree of polarization depends on the frequency $\w$ of the applied electric field, their complex permittivity $\epsilon(\w)$ and conductivity $\sigma(\w)$ of both, the particle and the suspending medium~\cite{Feldman2005}. It is well known that, the degree of polarization can be related with a relaxation time $\tau$. At the application of an external electric field, the polarization does not reach instantaneously its steady-state. Also, when the field is suddenly removed, a time $\tau$ elapses for the polarization to decay exponentially~\cite{raicu2015dielectric}. The overall electric behavior of particles in suspension is commonly analyzed by the effective medium theory and the Maxwell\textquotesingle s mixture model~\cite{Grimnes2002}. The aforementioned techniques relate the complex permittivity of the suspension with the permittivity of the particle, the suspending medium, as well as the volume fraction. Even though, these methods underestimate the relaxation theory behind the phenomenon under study. 

The classical technique for evaluating the electrical response of dielectric-conductive matter, is the so called electrical impedance spectroscopy (EIS), that relates the electrical properties with the frequency of the applied field. Thus, EIS has been attractive to experimentally measure the interaction of a time dependent electric field with a sample. An impedimetric measurement setup considers an applied voltage signal at an arbitrary frequency, and the measurement of a current flowing through the sample under study. The impedance $Z(j\w)\in\C$, is given by the ratio of the voltage and current, in the frequency domain. Typically, an impedance spectrum is extracted by a frequency sweep of a sinusoidal excitation ~\cite{Lasia2014}, and more recently using wide-band optimized signals~\cite{RAMIREZSYSID}. In practice, there are several issues to be tackled in order to avoid undesirable effects such as, the electrical double layer (EDL), and parasitic capacitance and inductance~\cite{Chassagne2016}. This, requires the design of adequate measurement electrodes~\cite{Chang2016}, or sophisticated instrumentation~\cite{Kortschot2014}.

Another challenging task is the impedance spectrum data interpretation, which requires carefully modelling but its physical understanding is not always obvious. This problem has been studied from different perspectives, for instance, by fitting the experimental data with parametric models such as the Cole-Cole model~\cite{Iglesias2017}, by rigorous theoretical modeling~\cite{Sun2007}, or by a lumped electrical equivalent circuit analysis~\cite{Sun2010}. Nevertheless, the aforementioned methods require prior knowledge of the composition, structure, and processes occurring within the sample under study.

In contrast, an alternative approach for EIS data interpretation  is the so called, distribution of relaxation times (DRT) model~\cite{Boukamp2015}. It is directly related with the relaxation theory, thus it can isolate all the relaxation processes from EIS spectrum, by transforming the data from the frequency domain to a time constant domain $\tau$. With knowledge of the DRT representation in terms of different relaxation processes comprised in a distribution function, one can build a suitable impedance model for further data evaluation. Despite of its narrow relationship with the electrical properties of colloids, to the best of our knowledge, the DRT model has not been applied in the analysis of EIS data taken from micro-particle suspensions.

We propose in this work a method for characterizing suspensions of micro-particles, with the main contributions of: (i) the introduction of a systematic procedure for estimating the concentration of dielectric particles suspended on a highly conductive matrix, (ii) by means of fast EIS measurements, and (iii) we present the first results towards the DRT model as an alternative approach to sense the concentration of micro-particles in a colloid, based on the relaxation theory.

The paper is organized as follows:
Section~\ref{Problem} formulates the problem, Section~\ref{Theo} introduces the useful concepts of impedance measurements and models for data analysis, Section~\ref{meas} presents the measurement procedure while experimental results are shown in Section~\ref{Results}, and Section~\ref{Conclu} presents the conclusions.
\section{Problem statement}\label{Problem}
Given a data set of impedance measured values $\{Z(j\w_1),Z(j\w_2),\dots, Z(j\w_N)\} \in \C$. Our goal is to introduce a method for systematically isolate the relaxation processes from the measured impedance spectrum, such that, a distribution function $\gamma(\ln \tau)$ will comprise the information of the involved processes in a time constant ($\tau$) domain. We propose that, with the retrieved function, one can obtain the concentration of the particles suspended in micro-colloids.
\section{Theoretical Background}\label{Theo}
This section attempts to introduce the basis of this work, for reader better understanding. 
\subsection{Electrical impedance spectroscopy}
Let us define the electrical impedance as the ratio of an applied voltage $V(\w_i)$ and a measured current $I(\w_i)$ at an arbitrary frequency $\w_i$, for $i=1,2,\dots,N$. The $i-$th impedance point is given by
\begin{equation}
\label{Z}
Z(j\w_i)\triangleq \frac{V(\w_i)}{I(\w_i)}=\abs{Z}\left(\cos(\phi) + j \sin(\phi)\right),
\end{equation}
where $j=\sqrt{-1}$, $\abs{Z}$ is the impedance magnitude, and $\phi$ is the phase angle between the voltage and current phasors. The impedance can be represented by its real $Z^\Re(\w)$ and imaginary $Z^\Im(\w)$ components.  The impedance spectrum is then formed by $N$ impedance data points, each one measured at an specific frequency.

\subsection{Equivalent electrical circuit for micro-colloids}
According to the literature, when measuring EIS in micro-colloids, there is a classical approach for interpreting the resultant spectra. The method is given in terms of an equivalent electrical circuit, whose parameters could be related with the composition and electrical processes within the sample. In Fig. 1, we show the equivalent circuit structure proposed in~\cite{Sun2010}. 
\begin{figure}[ht]
\centering
	   	 \def\svgwidth{0.25\textwidth}
	   	 \small
	   	 %% Creator: Inkscape inkscape 0.92.3, www.inkscape.org
%% PDF/EPS/PS + LaTeX output extension by Johan Engelen, 2010
%% Accompanies image file 'circ2.pdf' (pdf, eps, ps)
%%
%% To include the image in your LaTeX document, write
%%   \input{<filename>.pdf_tex}
%%  instead of
%%   \includegraphics{<filename>.pdf}
%% To scale the image, write
%%   \def\svgwidth{<desired width>}
%%   \input{<filename>.pdf_tex}
%%  instead of
%%   \includegraphics[width=<desired width>]{<filename>.pdf}
%%
%% Images with a different path to the parent latex file can
%% be accessed with the `import' package (which may need to be
%% installed) using
%%   \usepackage{import}
%% in the preamble, and then including the image with
%%   \import{<path to file>}{<filename>.pdf_tex}
%% Alternatively, one can specify
%%   \graphicspath{{<path to file>/}}
%% 
%% For more information, please see info/svg-inkscape on CTAN:
%%   http://tug.ctan.org/tex-archive/info/svg-inkscape
%%
\begingroup%
  \makeatletter%
  \providecommand\color[2][]{%
    \errmessage{(Inkscape) Color is used for the text in Inkscape, but the package 'color.sty' is not loaded}%
    \renewcommand\color[2][]{}%
  }%
  \providecommand\transparent[1]{%
    \errmessage{(Inkscape) Transparency is used (non-zero) for the text in Inkscape, but the package 'transparent.sty' is not loaded}%
    \renewcommand\transparent[1]{}%
  }%
  \providecommand\rotatebox[2]{#2}%
  \newcommand*\fsize{\dimexpr\f@size pt\relax}%
  \newcommand*\lineheight[1]{\fontsize{\fsize}{#1\fsize}\selectfont}%
  \ifx\svgwidth\undefined%
    \setlength{\unitlength}{113.52381321bp}%
    \ifx\svgscale\undefined%
      \relax%
    \else%
      \setlength{\unitlength}{\unitlength * \real{\svgscale}}%
    \fi%
  \else%
    \setlength{\unitlength}{\svgwidth}%
  \fi%
  \global\let\svgwidth\undefined%
  \global\let\svgscale\undefined%
  \makeatother%
  \begin{picture}(1,0.41318682)%
    \lineheight{1}%
    \setlength\tabcolsep{0pt}%
    \put(0,0){\includegraphics[width=\unitlength,page=1]{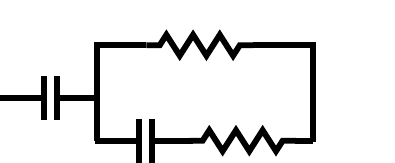}}%
    \put(0.09469664,0.25359739){\color[rgb]{0,0,0}\makebox(0,0)[lt]{\lineheight{1.25}\smash{\begin{tabular}[t]{l}$C_{\text{e}}$\end{tabular}}}}%
    \put(0.33253243,0.14789259){\color[rgb]{0,0,0}\makebox(0,0)[lt]{\lineheight{1.25}\smash{\begin{tabular}[t]{l}$C_{\text{p}}$\end{tabular}}}}%
    \put(0.57036822,0.14789259){\color[rgb]{0,0,0}\makebox(0,0)[lt]{\lineheight{1.25}\smash{\begin{tabular}[t]{l}$R_{\text{p}}$\end{tabular}}}}%
    \put(0.45145033,0.37251528){\color[rgb]{0,0,0}\makebox(0,0)[lt]{\lineheight{1.25}\smash{\begin{tabular}[t]{l}$R_{\text{m}}$\end{tabular}}}}%
  \end{picture}%
\endgroup%

	   	 \caption{Equivalent electrical circuit of a colloidal suspension of microparticles in a conductive medium.}
	   	 \label{fig:circ}
\end{figure}
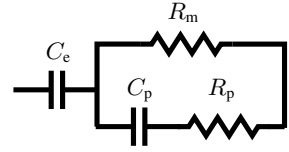

The impedance model for the circuit shown in Fig.~\ref{fig:circ} is parametrized in the vector $\btheta=\left[C_\text{e} \quad R_\text{m}  \quad R_\text{p} \quad C_\text{p}\right]$, and written as
\begin{align}
Z_{\text{cir}}(j\omega;\btheta)= \frac{1}{j\w C_\text{e}} + \frac{R_{\text{m}}(1+j\w R_\text{p}C_\text{p})}{j\w R_\text{m}C_\text{p} +(1+j\w R_\text{p}C_\text{p})},
\end{align}
where $C_\text{e}$ represents the capacitance of the electrodes, $R_{\text{m}}$ is the medium resistance, $R_\text{p}$ corresponds to the resistance of the dielectric particles, and $C_\text{p}$ is their capacitance. 

The main disadvantage of the circuit direct analysis is that the frequency measurement range needs to be high enough in order to diminish the electrodes capacitance, or the instrumentation needs to have high sensitivity.

\subsection{Distribution of relaxation times model}
\begin{figure}[h!]
\centering
	   	 \def\svgwidth{0.45\textwidth}
	   	 \small
	   	 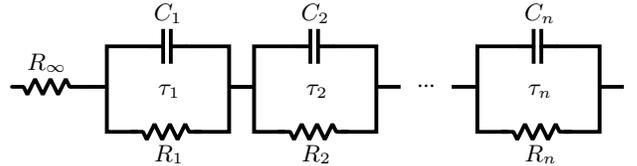
	   	 \caption{Equivalent electrical circuit of the distribution of relaxation times (DRT) model.}
	   	 \label{fig:rand1}
\end{figure}

The distribution of relaxation times (DRT) is an alternative approach to analyze EIS measurements, commonly used in electrochemical applications. By using DRT it is possible to obtain a distribution function that joins up the polarization and conduction processes involved in the sample under study. The procedure is done by fitting the measured EIS data to the impedance model given by
\begin{equation} \label{eq:ZDRT}
Z_{\text{DRT}}(j\omega)=R_{\infty}+\int_0^\infty \frac{\gamma(\ln\tau)}{1+j\omega\tau}d(\ln\tau),
\end{equation}
where $R_{\infty}$ is named high-frequency resistance and $\gamma(\ln\tau)$ is the distribution function. Solving~(\ref{eq:ZDRT}) for this latter is an inverse problem with an infinite number of solutions. In a previous work~\cite{RAMIREZSYSID}, we proposed a parametric kernel method for solving it using a regularized least-squares algorithm. Once the $\gamma(\cdot)$ function is retrieved, it can be seen as a time constant domain spectrum, transformed from the frequency domain. By means of DRT, different relaxation mechanisms can be easily isolated and identified in terms of the equivalent electrical circuit shown in Fig~\ref{fig:rand1}. 

The DRT equivalent circuit comprises a series connection of $n$ (theoretically $n=\infty$) $RC$  parallel networks, each one related with a polarization process characterized by an $i-$th relaxation time and defined as, $\tau_i=R_iC_i$, for $i=1,\dots,n$. Each characteristic time can be easily identified by means of local maxima points of the distribution function, in a plot of $\gamma(\cdot)$ against $\tau$. Thus, one can state that the number of maxima is equal to the number $n$ in the circuit of Fig.~\ref{fig:rand1}.

\section{Measurement procedure}\label{meas}
\begin{figure}[htb]
\centering
	   	 \def\svgwidth{0.5\textwidth}
	   	 \small
	   	 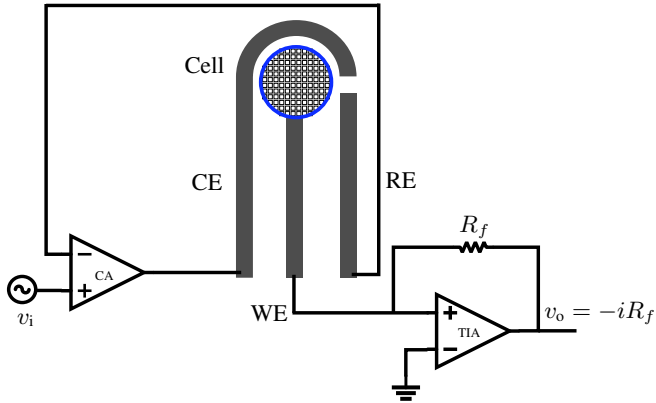
	   	 \caption{Schematic diagram of the potentiostat circuit used for EIS measurements.}
	   	 \label{fig:poten}
\end{figure}

For EIS measurements, we use a custom impedance analyzer designed by our group of work. It permits to perform fast and accurate impedance measurements by means of a broadband multisine signal and an unbiased non-parametric estimator. For this work, we use an electrochemical cell in a potentiostat configuration as shown in Fig.~\ref{fig:poten}. 

\subsection{Potentiostat and electrochemicall cell}
The electrochemical cell has three electrodes made of platinum (Pt) over a glass substrate, fabricated using optical lithography. 
The sensing element is a micro-electrode array (MEA), in a 3 mm diameter surface with 620 micro-electrodes of $10\mu\text{m}$, and a distance of $10\mu\text{m}$ among them. In the electrochemical setup the MEA acts as the working electrode (WE). Herein, a multisine voltage signal ($v_{\text{i}}$) is applied to the counter electrode (CE) by means of a control amplifier (CA), which compensates its output by measuring the voltage drop in the reference electrode (RE). The current flowing through the sample placed in the WE is measured using a transimpedance amplifier (TIA), whose output voltage ($v_{\text{o}}$) is proportional to the current $i$, by the feedback resistor $R_f$. 
\subsection{Non-parametric impedance measurement estimation}\label{nonparam}
The applied voltage $v_{\text{i}}$ has the structure of a random phase multisine signal given by
\begin{align}
\label{msine}
v_{\text{i}}(t)=\sum_{k=1}^{F} A_k\cos(\omega_kt+\delta_k), 
\end{align}
where $A_k$ is the amplitude, $\delta_k$ is the phase uniformly distributed in the range $\mathrm{(0,2\pi]}$, and $\omega_k=2\pi f_k$ is $k-$th excited frequency, for $k=1,2,\dots,F$, with $F=52$ for this work. The signal has an amplitude of $\pm 0.1 \text{V}$, in the range from 1 kHz to 1 MHz.  

For accurate impedance measurements, we acquire an integer number $M$ of periods of the input $v_{i}(t)$ and output $v_{o}(t)\propto i(t)$ signals, whose are then transformed to the frequency domain by the discrete Fourier transform (DFT). The input/output spectra are defined as $V_{\text{i}}(k)\triangleq \Fsp \{v_{\text{i}}(t)\}$ and $V_{\text{o}}(k)\triangleq \Fsp \{v_{\text{o}}(t)\}$, respectively, with $\Fsp\{\cdot\}$ denoting the DFT operator. Thus, an unbiased estimator for the measured impedance $\hat{Z}(j\w)$, is given as the ratio of the DFT sample means over $p=1,2,\dots,P$ signal periods,
\begin{align}
\hat{Z}(j\w)\triangleq
%\frac{\tilde{U}(k)}{\tilde{Y}(k)}=
\frac{\frac{1}{M}\sum_{m=1}^{M} V_{\text{i}}^{[m]}(k)}{\frac{1}{M}\sum_{m=1}^{M} V_{\text{o}}^{[m]}(k)/R_f}.\label{eq:estim}
\end{align}
The estimator increases the signal-to-noise ratio (SNR) by measuring a large number of periods, by the properties of the spectral averaging technique~\cite{Pintelon2001}.  
Therefore, we can obtain a non-parametric estimate such that $\hat{Z}(j\w) \approx Z(j\w)$
Finally, the measured impedance spectrum $\hat{Z}(j\w)$ is fitted to the DRT model in~(\ref{eq:ZDRT}), thus giving a distribution function $\gamma(\ln\tau)$, as a time constant domain representation for each measured sample.

\section{Experimental Results}\label{Results}
\subsection{Sample preparation}
The micro-colloid samples are composed by $\text{SiO}_2$ micro-spheres with a diameter of $d=0.8\mu\text{m}$. These are suspended on a conductive phosphate buffer saline (PBS) matrix at four different concentrations $\kappa$ in weight percentage (wt.\%), see Table~\ref{tab1}. According to the literature, the $\text{SiO}_2$ particles exhibits a large dielectric behavior, with permittivity  $\epsilon_{\text{SiO}_2}= 4.5$ and conductivity $\sigma_{\text{SiO}_2}=  10^{-8} \, \text{S/m}$, while the PBS is highly conductive, with $\epsilon_{\text{PBS}}= 78$, and $\sigma_{\text{PBS}}= 0.15\, \text{S/m}$. 

\begin{figure}[H]
\centering
	   	 \def\svgwidth{0.45\textwidth}
	   	 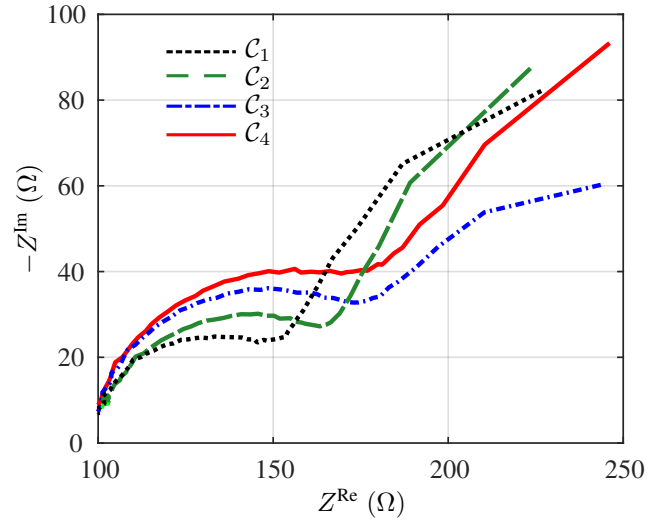
	   	 \caption{Nyquist diagram for the impedance measurements, for the micro-colloidal samples of $\text{SiO}_2$ particles in a PBS suspending matrix.}
	   	 \label{fig:nyq_sio2}
\end{figure}

\subsection{EIS measurements}
Using the non-parametric impedance estimator introduced in Section~\ref{nonparam}, we measured the four samples using the potentiostat circuit, in which 200$\mu$L of the micro-colloids were placed on the working electrode. We take $P=8$ periods of the multisine voltage and current signals, for estimating the impedance $\hat{Z}$ in (\ref{eq:estim}). As a result, we obtained the impedance data in its real $Z^\Re$ and imaginary $Z^\Im$ components. In Fig.~\ref{fig:nyq_sio2} we depict the plots of $\hat{Z}$ for the evaluated concentrations, in a Nyquist diagram. 

From Fig.~\ref{fig:nyq_sio2}, one can see two main processes involved in measurements. In the low frequencies regime, a straight-line-kind trend appears, which is due to the interaction of the WE with the sample~\cite{Barbero2008}. Meanwhile, at high frequencies, one can see a semi-circle, which suggests the impedance signal related with the sample under study. It is worth to notice that the four plots tend to around 100 $\Omega$ in the real component, thus one can relate this value with the resistance of PBS liquid matrix. Ultimately, by this inspection it is difficult to directly extract a relationship of the concentrations with the relaxation theory, thus an alternative approach will be introduced in the next section.
  
\subsection{DRT analysis}
The measured EIS data were analyzed using the DRT model. Therefore, we obtained four different distribution functions $\gamma(\cdot)$, each one associated to concentration $\Csp_k$, for $k=1,2,3,4$.

\begin{figure*}[t]
\centering
	   	 \def\svgwidth{0.85\textwidth}
	   	 \small
	   	 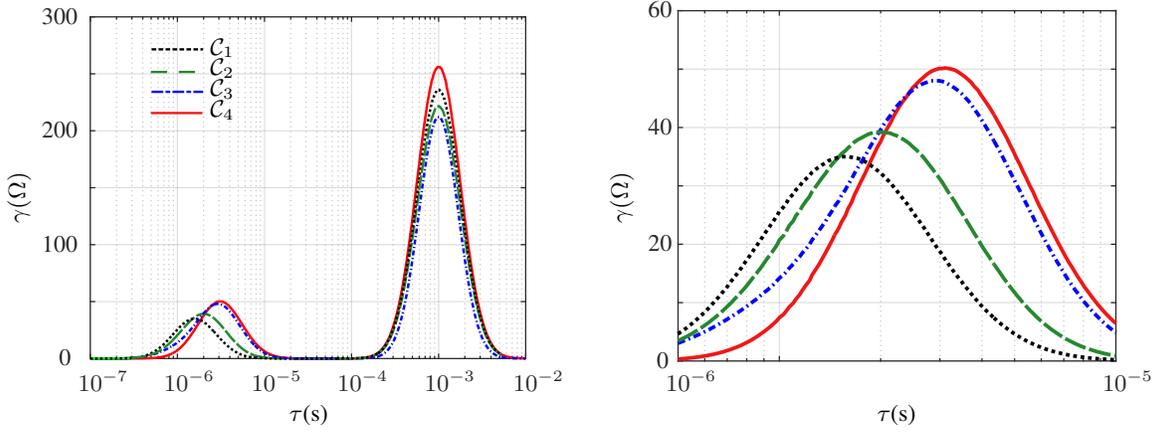
	   	 \caption{Distribution function $\gamma(\cdot)$ in the time constant domain $\tau$. (a) For the micro-colloidal samples of $\text{SiO}_2$ particles suspended in PBS at four different concentrations. (b) Zoom of the distribution function for a time constant scale in the range from $\tau \approx 10^{-7}$ to $10^{-5}$.}
	   	 \label{fig:drt_sio2}
	   	 
\end{figure*}

Fig.~\ref{fig:drt_sio2}(a) shows the retrieved $\gamma(\cdot)$  function measured in $\Omega$, in a time constant domain $\tau$. The curves exhibit two local maxima points, for each measured sample. 
It is worth to notice that, in the region of small relaxation times, i.e. $\tau \approx10^{-6}$ s, the presence of a Gaussian-like peaks in the distribution functions are well differentiated among them, in terms of its amplitude and location in the $\tau$ axis.
Nevertheless, for high relaxation time values around $\tau\approx 10^{-3}$, one can see that the four maxima are very close in both amplitude and location. We can expect two situations from the DRT analysis:
\begin{itemize}
\item At small relaxation times, the distribution function $\gamma(\cdot)$ is giving information related with the concentration of micro-particles in the colloid.

\item At high relaxation times, dominant and consistent presence of local maxima presumes the effect of a double layer capacitance or Warburg impdance due to the electrode-electrolyte interface~\cite{Chassagne2016}. 
\end{itemize}

Under these conditions, the further analysis will be focused on the small relaxation times region, where likely, the useful information of the micro-colloid could be obtained. Fig.~\ref{fig:drt_sio2}(b) shows a zoom of the distribution functions in a time constant range from $10^{-7}$ to $10^{-5}$ seconds. In this plot, one can notice a shift of the maximum as the concentration increases, thus, the maxima are associated to a characteristic relaxation time. And it make sense, since the presence of a large number of dielectric-insulating particles within the colloid, can be associated to a higher effective resistance of the sample, which is proportional to the relaxation time by $\tau=RC$.

Table~\ref{tab1} summarizes the evaluated samples and the relaxation times obtained with the DRT model from the EIS measurements shown in Fig.~\ref{fig:nyq_sio2}.
\begin{table}[H]
\caption{Micro-colloid concentrations in weight and retrieved relaxation times.}
\begin{center}
\begin{tabular}{|c|c|c|}
\hline
\textbf{Sample} & $\kappa$~(wt.\%)& $\mathbf{\tau}\times 10 ^{-6}$~(s)\\
\hline
$\Csp_1$& 0.1& 1.60 \\
\hline
$\Csp_2$& 0.5&  2.00\\
\hline
$\Csp_3$& 1.0 & 2.85 \\
\hline
$\Csp_4$ & 1.5& 3.33 \\
\hline
\end{tabular}
\label{tab1}
\end{center}
\end{table}

\subsection{Sensor calibration}
\begin{figure}[t]
\centering
	   	 \def\svgwidth{0.45\textwidth}
	   	 \small
	   	 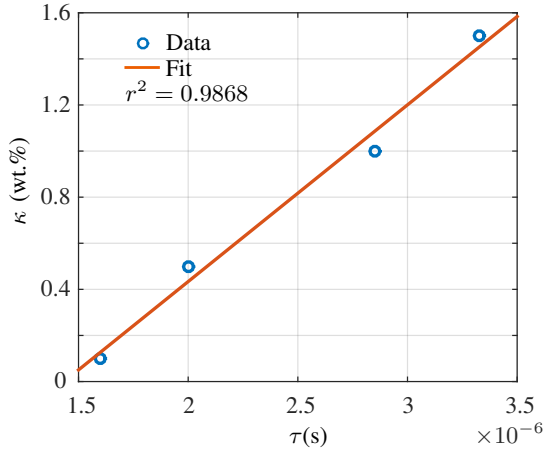
	   	 \caption{Calibration curve of the micro-particles concentration $\kappa$ of the micro-colloids, as a function of the retrieved values of the relaxation time $\tau$.}
	   	 \label{fig:concentra_1}
\end{figure}

Using the aforementioned analysis, the proposal of an EIS-DRT sensing method is given by the retrieved relaxation times and the concentration of the particles. In Fig.~\ref{fig:concentra_1} we depict the sensor calibration curve, where one can see the relaxation times  $\tau$ as a function of the concentration in weight (wt.\%) for the four samples.

In the calibration curve one can see the experimental points (circles) and the model that best fits them (straight line), which obeys a linear trend. Thus, the model of the proposed sensor is given by
\begin{align}
\kappa(\tau;\btheta)=\theta(a)\;\tau+\theta(b),
\end{align}
parametrized in $\btheta=\left[a\;b\right]^\top$, being $a$ the slope, and $b$ the intercept of the linear model. The optimal straight line $\kappa^\star(\cdot)$, in a least-squares sense, is given by the estimates $\hat{a}$ and $\hat{b}$. Finally, the goodness of the fit is given by the correlation coefficient $r^2=0.9868$.

Using the optimal model $\kappa^\star$, the sensor sensitivity $\Ssp$ is straightforward computed as, 

\begin{align}
\Ssp \triangleq \frac{\partial \kappa^\star}{\partial \tau} =0.76 \left(\frac{\text{wt.\%}}{\mu\text{s}}\right).
\end{align}

Whereas the intercept $\hat{b}=0.05\,\text{wt.\%}$, can be related with the detection limit. In this sense, the measurement resolution in $\tau$, is thus determined by the frequency resolution of EIS spectrum, which in our case is $1.59\times 10^{-7}$ s. That conducts to a resolution in the concentration measurement of $0.01\,\text{wt.\%}$. Ultimately, a deeper study could relate the retrieved parameters from the DRT-based analysis with the intrinsic properties of the micro-colloids, in terms of effective permittivity and conductivity. 

\section{Conclusion}\label{Conclu}
We propose an alternative method towards sensing the concentration of dielectric particles in suspension. The method is based upon the relaxation theory applied to spectral impedance measurements, by using the distribution of relaxation times model. Results showed that the EIS-DRT method, fairly represents the relaxation processes involved in micro-colloidal samples. In contrast to the classical methods, our proposal does not require prior knowledge of the structure and composition of the sample under study, and can be systematically performed. The EIS-DRT-based concentration sensing method exhibits high sensitivity and resolution, besides a narrow relationship with physical interpretation. One of the main interest on measuring colloidal samples is due to the fact that they can mimic a biological cell suspension. Therefore, we expect that the proposal could be applied for further studies in biomedical or clinical fields.  

\section*{Acknowledgment}
This work was partially supported by UNAM-PAPIIT IT100518 grant, and RGRCH CONACYT PhD studies grant~(CVU 555791).

\section*{References}
\bibliographystyle{IEEEtran}

\end{document}